\documentclass[%
 prl
 aip,
 jmp,%
 amsmath,amssymb,
twocolumn,superscriptaddress]{revtex4-2}

\usepackage{graphicx} 
\usepackage{hyperref} 
\usepackage{xcolor}   
\usepackage{physics}  
\usepackage{diagbox}  
\usepackage{dcolumn}  
\usepackage{booktabs} 
\usepackage{makecell} 
\usepackage[english]{babel} 
\usepackage{multirow}

\hypersetup{
    colorlinks=true,
    linkcolor=blue,
    citecolor=blue,
    filecolor=blue,
    urlcolor=blue
}

\newcommand{\UMD}{University of Maryland, College Park, Maryland, USA}

\newcommand{\NIST}{Sensor Science Division, National Institute of Standards and Technology, Gaithersburg, Maryland 20899, USA}

\newcommand{\RUG}{
Van Swinderen Institute for Particle Physics and Gravity,
\\University of Groningen, Nijenborgh 4, 9747AG Groningen, The Netherlands\\
}

\newcommand{\CUB}{Department of Physical and Theoretical Chemistry, Faculty of Natural Sciences, 
Comenius University, Mlynsk\'a dolina, 84215, Bratislava, Slovakia}

\begin{document}

\title{Radiative Decay Rate and Branching Fractions of MgF}

\date{\today}

\author{E. B. Norrgard$^*$}
 \affiliation{\NIST}
 \email{eric.norrgard@nist.gov}

 \author{Yuly Chamorro}
\affiliation{\RUG}

\author{C. C.Cooksey}
 \affiliation{\NIST}

 \author{S. P. Eckel}
 \affiliation{\NIST}

\author{N. H. Pilgram}
\affiliation{\NIST}
\author{K. J. Rodriguez}
\affiliation{\NIST}
\affiliation{\UMD} 

\author{H. W. Yoon}
\affiliation{\NIST}

\author{Luk\'a\v{s} F. Pa\v{s}teka}
\affiliation{\RUG}
\affiliation{\CUB}

\author{Anastasia Borschevsky}
\affiliation{\RUG}

\begin{abstract}
We report measured and calculated values of radiative decay rates and vibrational branching fractions for the A$^2\Pi$ state of MgF.
The decay rate measurements use time-correlated single photon counting with roughly 1\,\% total uncertainty.
Branching-fraction measurements are performed using two calibrated imaging systems to achieve few percent total uncertainty.
We use the highly accurate multireference relativistic \textit{ab initio} methods to calculate the Franck-Condon factors and transition dipole moments required to determine the decay rates and the branching fractions. The measurements provide a precision benchmark for testing the accuracy of the molecular structure calculations. The determination of the decay rate and vibrational branching fractions can be used to inform future optical cycling and laser cooling schemes for the MgF molecule.

\end{abstract}

\maketitle

\section{Introduction}
Recently, magneto-optical traps containing a record number $N\approx 10^5$ trapped molecules have been demonstrated for CaF \cite{Anderegg2018} and YO \cite{Ding2020}.
  Further increases to the number of molecules trapped are desireable for several laser-cooled molecule applications. Studies of cold collisions and cold chemistry \cite{Tomza2015,Liu2020,Jurgilas2021} require high reaction rates, motivating a large $N$ to achieve a high molecule density.  Because signal-to-noise ratios generally are proportional to $\sqrt{N}$, molecule-based sensors \cite{Alyabyshev2012,Norrgard2021} directly benefit. Other applications, such as quantum information and quantum simulation \cite{DeMille2002,Moses2016,Blackmore2019}, require low temperature $T$ rather than large $N$. Because the ultimate step in cooling is often evaporation \cite{Son2020}, which sacrifices $N$ to achieve lower $T$, such applications also benefit from a large initial $N$. 

Efficient slowing of  molecular beams is vital to achieve a large trapped molecule number $N$.  
A large deceleration $a$ from laser slowing \cite{Barry2012} is desired because the distance required to stop molecules $d \propto 1/\sqrt{a}$
Additionally, as molecules are slowed, the molecular beam brightness falls as $1/d^2$ if the beam's transverse velocity distribution is not modified.
Thus, simple scaling arguments show that the trapped molecule number $N$, which is proportional to beam brightness, is proportional to $a$.
Simultaneously applying transverse radiative cooling and longitudinal radiative slowing often does not help; near saturation, the two processes compete with each other, leading to a decrease in trappable beam brightness \cite{DeMille2013}.
The maximum possible deceleration is $a_{\rm{max}}=h \Gamma /(2 m  \lambda)$, so large deceleration is achieved with low mass $m$, fast radiative decay rate $\Gamma$, and short wavelength $\lambda$.  All three of these factors favor MgF over current state-of-the-art CaF and YO experiments, with expected trapping improved by as much as a factor of $7$ or $20$, respectively.

Laser slowing a MgF beam to a stop from an initial velocity of $150$\,m/s, typical of a cryogenic beam, requires optical cycling of a number of photons $N_\gamma \approx 6000$.  At a minimum, a preliminary attempt to laser cool MgF requires knowledge of all vibrational branching fractions $b_{0v^{\prime\prime}} > 1/N_\gamma \approx 1.7\times 10^{-4}$, as each corresponding ground vibrational state $v^{\prime\prime}$ will need to be addressed by a laser.
Like other alkaline earth monofluorides, MgF is expected to have a nearly diagonal array of Franck-Condon factors, enabling laser cooling with few vibrational state repump lasers \cite{DiRosa2004}.
But despite spectroscopic studies of MgF dating back nearly 100 years \cite{Jevons1929,Jenkins1934}, we are not aware of any experimental determinations of vibrational branching at the $10^{-4}$ level.  Moreover, while MgF is a relatively simple system where relativistic and correlation effects are minimal, predictions of the number of $b_{0v^{\prime\prime}} \gtrsim  10^{-4}$ are variously two \cite{Pelegrini2005,Xu2016}, three \cite{SINGH1969}, or four \cite{Kang2015} in the literature.  An accurate measurement of the vibrational branching fractions $b_{0v^{\prime\prime}}$ for MgF is therefore needed to inform allocation of laser resources for laser cooling experiments, as well as to benchmark calculation methods for their applicability to more complicated laser coolable molecules (e.g. molecules which contain heavy nuclei and/or more than two atoms).

An upper limit on the spontaneous decay rate $\Gamma$ of the MgF  A$^2\Pi_{1/2}$ state was recently reported with approximately 4\,\% fractional uncertainty in Ref.\, \cite{Doppelbauer2022}.  There, $\Gamma$ was determined by measuring  laser induced fluorescence as a function of laser frequency and extrapolating the fitted Lorenzatian linewidth to zero power.  Several groups have performed calculations of $\Gamma$ \cite{Pelegrini2005, Kang2015}, roughly consistent with the  measurement of Ref.\,\cite{Doppelbauer2022}.

Here we report measured values of radiative decay rates of the A$^2\Pi$ state and vibrational branching fractions of the A$^2\Pi_{1/2}\rightarrow {\rm X}^2\Sigma^+$  transition in the MgF molecule.
Branching fractions are measured by exciting the transition of relevance for laser cooling, which is P$_1(1)$/Q$_{12}(1)$ \cite{Stuhl2008}.
We find good agreement with theory, which is described in Sec.~\ref{sec:theory}.
Our experimental results are detailed in Secs~\ref{sec:branching_fractions} and~\ref{sec:decay_rates} for the branching fractions and decay rates, respectively.
Finally, we conclude with remarks in Sec.~\ref{sec:conclusion}.

\section{Theory}\label{sec:theory}

\begin{figure}[b]
    \centering
    \includegraphics[scale=0.45]{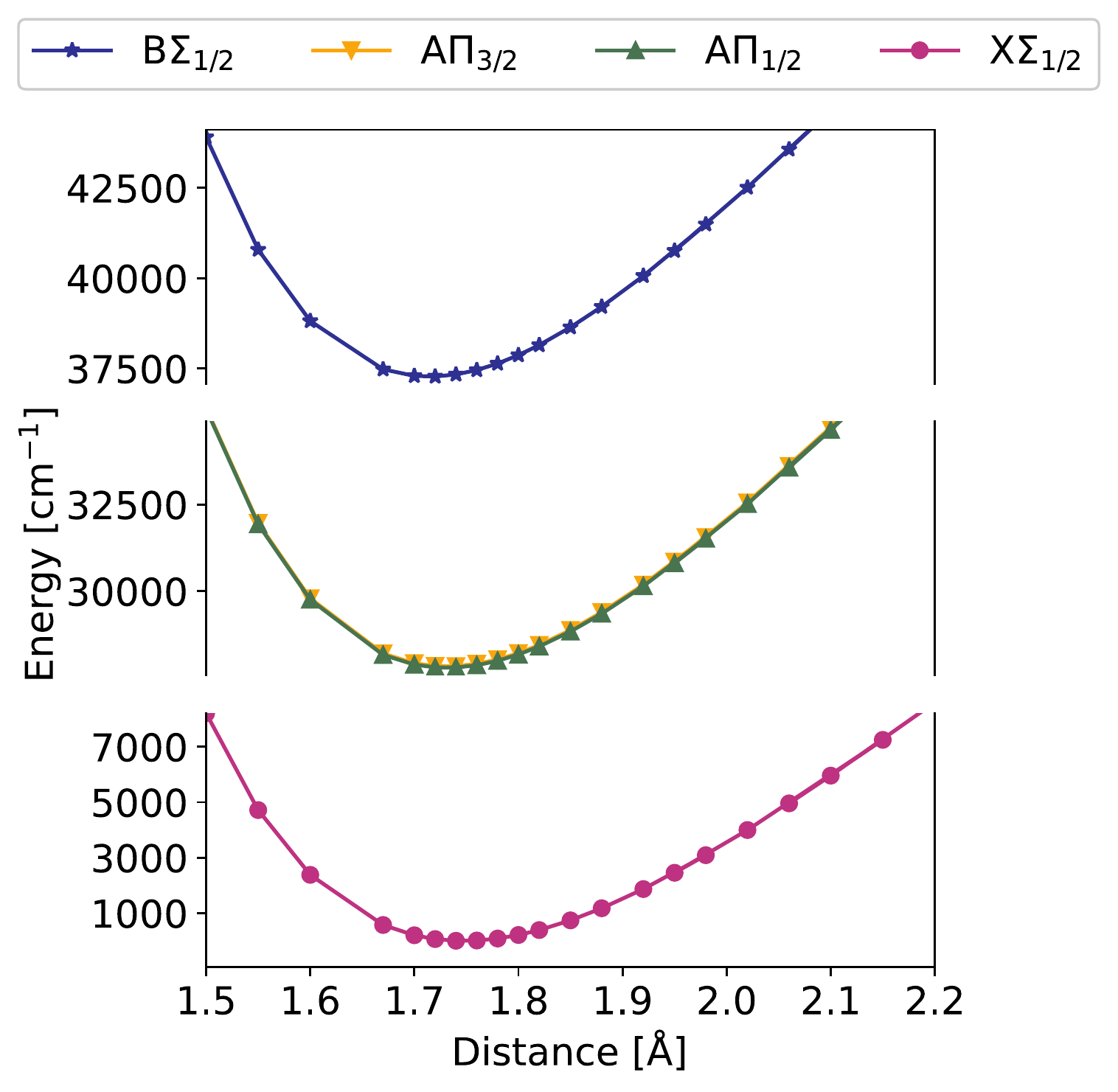}
    \caption{PECs for the ground and low-lying excited states of MgF obtained at 4c-FSCC(0,1) level and using the d-aug-dyall.ae4z basis set.}
    \label{fig:PES}
\end{figure}

The X$^2\Sigma_{1/2}$ ground state and the A$^2\Pi_{1/2}$, A$^2\Pi_{3/2}$, and B$^2\Sigma_{1/2}$ excited states of the MgF molecule were investigated using the relativistic 4-component Fock-space coupled cluster method (4c-FSCC), following a similar approach to Ref.\,\cite{HaoPasViss19}. We initially solve the coupled cluster equations for the closed-shell molecular ion, MgF$^+$. This serves as a reference for the subsequent FSCC(0,1) calculation of ground and excited states of the neutral MgF. We employed the dyall.d-aug-ae4z \cite{Dya16} basis set and the default settings in the DIRAC19 program \cite{SauBasGom20,DIRAC19}. 
We use the constructed potential energy curves (PECs) shown in Fig.~\ref{fig:PES} and the Twofit utility program in DIRAC to calculate the spectroscopic constants, as defined by Herzberg \cite{Her13}. $T_e$ represents the minimum on the electronic state PEC (relative to the ground state). Table \ref{tab:spectroscopic} shows that the obtained spectroscopic constants are in an excellent agreement with the previously reported experimental values.

\begin{figure}[b]
    \centering
    \includegraphics[width=0.48\textwidth]{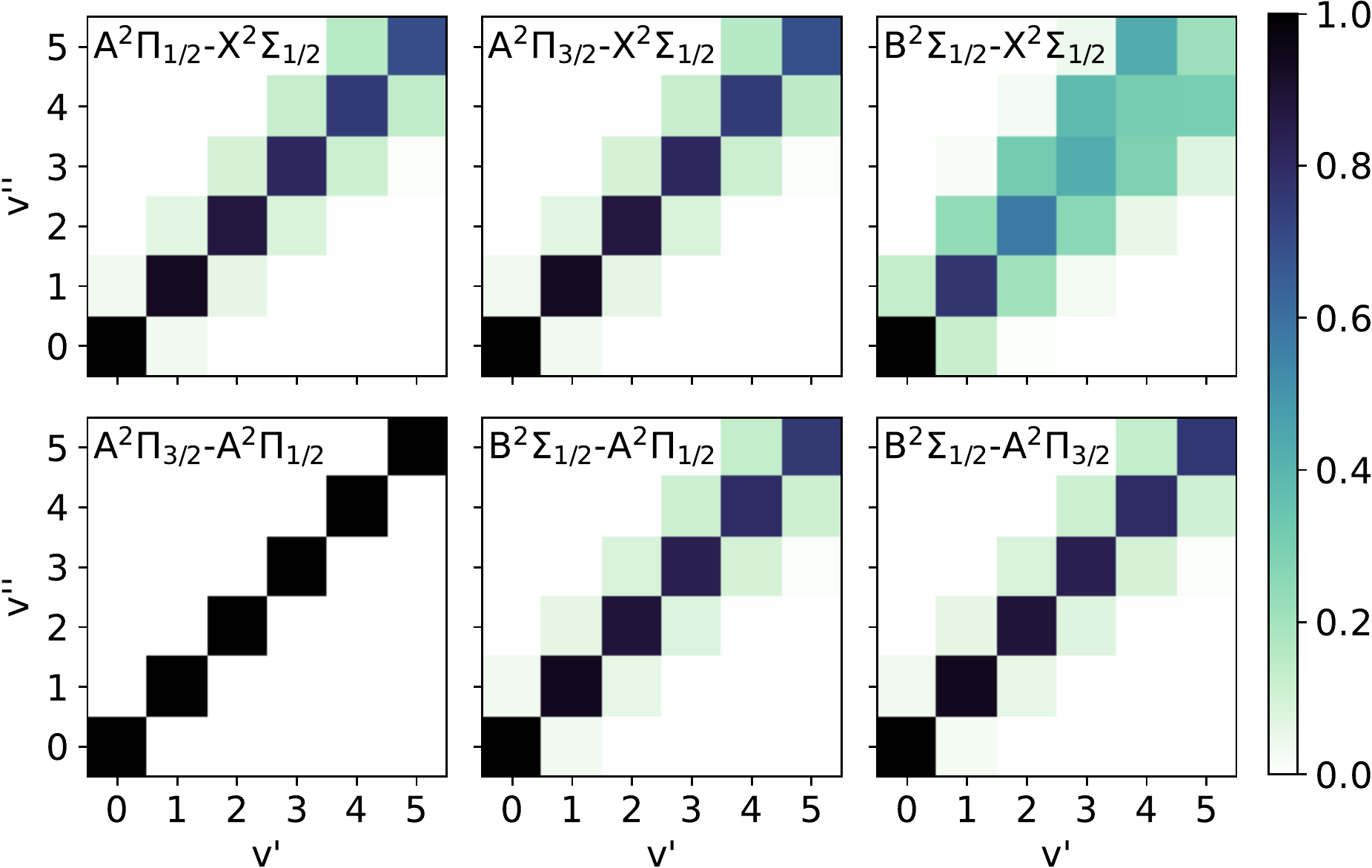}
    \caption{FCFs for the vibronic transitions between the upper ($v'$) and lower ($v''$) electronic states. Table \ref{tab:FCF} in SI presents the corresponding numerical values.}  
    \label{fig:FCFs}
\end{figure}

\begin{table*}[]
\centering
\begin{tabular}{lllllll}
\hline \hline
 & $R_e$ & $\omega_e$  & $\omega_e\chi_e$  & $B_e$ & $T_e$  & Reference\\
 \hline
X$^2\Sigma_{1/2}$ & 1.747 & 729.1 & 4.23 & 0.519 &  & This work\\
                  & 1.749937(1)& 720.14042(30) & 4.26018(16) & 0.519272510(42) & & \cite{BarZhaGuo95}\\
                  & 1.7500 & 721.6  & 4.94 & 0.51922 &  &\cite{BarBea67}$^{\star}$\\
A$^2\Pi_{1/2}$    & 1.731 & 757.2 & 4.27 & 0.529 & 27794 & This work \\
                  & 1.7469 & 747.999(1) & 4.274(5) & 0.52106(6) & 27815.646(1)$^{\ddagger}$ & \cite{GuXiaYu22}\\ 
                  & 1.7470 & 746.6 & 3.25 & 0.52105 & 27798.5$^{\dagger}$ & \cite{BarBea67}$^{\star}$\\
A$^2\Pi_{3/2}$ & 1.731 & 757.4 & 4.27 & 0.529 & 27832 & This work \\
                & 1.7212 & 748.097(6) & 4.273(2) & 0.53676(5) & 27852.005(5)$^{\ddagger}$ & \cite{GuXiaYu22}\\
                & 1.7470 & 746.6 & 3.25 & 0.52105 & 27834.9$^{\dagger}$ & \cite{BarBea67}$^{\star}$\\
B$^2\Sigma_{1/2}$ & 1.714 & 775.7 & 4.98 & 0.539 & 37277 & This work\\
 &1.7185 & 762.2 & 5.63 & 0.53844 & 37167.3 & \cite{BarBea67}$^{\star}$\\
\hline
\end{tabular}
\caption{Equilibrium bond lengths in \AA \ and spectroscopic constants in cm$^{-1}$ of the ground and the low-lying excited states of MgF obtained at the 4c-FSCC(0,1) level of theory and using the d-aug-dyall.ae4z basis set. $^{\star}$Constants calculated from the reported $B_v$ and $T_v$ values for $v=0,1$.  $^{\ddagger}T_e$ values originally reported in \cite{GuXiaYu22} were equated with the band origins, and are thus corrected here using the other reported spectroscopic constants for the sake of consistency with the values in this work and Ref.~\cite{BarBea67}. $^{\dagger}$Obtained from the reported mean $\Lambda$-S state and the regular spin-orbit coupling.}  
\label{tab:spectroscopic}
\end{table*}

The transition rate between an upper ($n' v'$) and a lower ($n'' v''$) vibronic state, $\Gamma_{n'v'n''v''}$, is defined in terms of the transition dipole moment (TDM) function $M_{n'n''}(R)$ for the $n'$ and $n''$ electronic states, and the transition frequency  $\omega_{n'v'n''v''}$ between the $n'v'$ and $n''v''$ states, 

\begin{equation}\label{eq:transisiton rate}
    \Gamma_{n'v'n''v''}=\frac{\omega^3_{n'v'n''v''}}{3\pi\hbar\epsilon_0 c^3} |\bra{v'}M(R)_{n'n''}\ket{v''}|^2,
\end{equation}
where $\hbar$ is the reduced Planck constant and $\epsilon_0$ the permittivity of the vacuum.  
In the $r$-centroid approximation \cite{NicJar56}, the above integral can be expressed as the product of the constant square of the TDM  $M_{n'n''}^2$ and the square of the overlap of the vibrational wave functions, $q_{v^\prime v^{\prime\prime}}\,=\,|\bra{v'}\ket{v''}|^2$, also called the Franck--Condon factors (FCFs),\\

\begin{equation}\label{eq:r-centroid}
       \Gamma_{n'v'n''v''}=\frac{\omega^3_{n'v'n''v''}}{3\pi\hbar\epsilon_0 c^3} q_{v^\prime v^{\prime\prime}} M_{n'n''}^2.
\end{equation}
Figure \ref{fig:FCFs} presents the FCFs $ q_{v^\prime v^{\prime\prime}}$ between vibrational levels up to $v=5$ of the ground and excited states of MgF extracted from the LEVEL16 program \cite{Roy17} (corresponding numerical values are collected in Table \ref{tab:FCF} in SI).

Table \ref{tab:TDM} presents the TDMs $M_{n'n''}$ for the first four electronic states of MgF. TDMs were obtained using the multireference configuration interactions level of theory (MRCI) in the MOLPRO program \cite{MOLPRO_brief}. In the MRCI approach, we used the complete active space (9,9) reference allowing the single unpaired electron to occupy the Mg $3s, 3p, 4s$ orbitals and the F $2s, 2p$ orbitals. The same basis set as the one employed for calculating the PECs, i.e.\ the dyall.d-aug-ae4z basis set \cite{Dya16} was used and all electrons were correlated. We calculate the TDMs using the MRCI method as this is not yet implemented in the FSCC method. To verify the validity of our approach, we compare the dipole moments calculated using FSCC and MRCI (see discussion in the SI). We found no significant differences between the different methods and very good agreement with the experimental values when available.

The decay rate $\Gamma_{n'v'}$ (or inverse of the lifetime $\tau_{n^{\prime}v^{\prime}}$) of an excited state $(n^{\prime}v^{\prime})$ may be expressed as the summation of Eq.\,\eqref{eq:transisiton rate} over all lower states $(n''v'')$
\begin{equation}
   \Gamma_{n'v'}  =\frac{1}{\tau_{n^{\prime}v^{\prime}}} =\sum_{n^{\prime\prime} v^{\prime\prime}}\Gamma_{n'v'n''v''},
\end{equation}
The lifetimes were calculated using the LEVEL program \cite{Roy17}, and are presented in Table \ref{tab:lifetime}. There is very good agreement between values determined by the full $R$-dependent TDM calculation and the r-centroid approximation. 
 Thus the r-centroid approximation \eqref{eq:r-centroid} is easily applicable for this system in future studies. Decay rates are compared to experiment later in the paper in Table \ref{tab:decay rate comparison}.

The branching fractions $b_{v^\prime v^{\prime\prime}}$ from a given excited state ($n^\prime v^\prime)$ is expressed as
\begin{equation}
    b_{n^\prime v^\prime n^{\prime\prime} v^{\prime\prime}} = \frac{\Gamma_{n'v'n''v''}}{\Gamma_{n'v'}}.
\end{equation}
In the $r$-centroid approximation, the branching fractions for the A$^2\Pi_{1/2}$ state are simply the the FCFs renormalized with an $\omega^3_{n'v'n''v''}$ weighting.  

 In the remainder of this work, we consider only transitions between the A$^2\Pi$ and X$^2\Sigma^+$ states, and labels $n^\prime$ and $n^{\prime\prime}$ are suppressed for simplicity of notation.

\begin{table}[]
\centering
\begin{tabular}{ll}
\hline\hline
Transition & TDM\\
\hline
 A$^2\Pi_{1/2}$ -- X$^2\Sigma_{1/2}$ & 1.802\\
 A$^2\Pi_{3/2}$ -- X$^2\Sigma_{1/2}$ & 1.802\\
 B$^2\Sigma_{1/2}$ -- X$^2\Sigma_{1/2}$& 1.492\\
 A$^2\Pi_{3/2}$ -- A$^2\Pi_{1/2}$& 0.001 \\
 B$^2\Sigma_{1/2}$ -- A$^2\Pi_{1/2}$&0.504\\
 B$^2\Sigma_{1/2}$ -- A$^2\Pi_{3/2}$&0.504\\
\hline \hline      
\end{tabular}
\caption{Calculated transition dipole moments 
(in $e a_0$, where $e$ is the elctron charge and $a_0$ is the Bohr radius) for the ground state bond length (1.747 \AA) using the SO-MRCI method and the d-aug-dyall.ae4z basis set.}
\label{tab:TDM}
\end{table}

\begin{table}[]
\centering
\begin{tabular}{lll}
\hline
\hline
State & $\tau$ [Eq.\,\eqref{eq:r-centroid}] & $\tau$ [Eq.\,\eqref{eq:transisiton rate}]\\
\hline
 A$^2\Pi_{1/2}$ & 7.09 & 7.08\\
 A$^2\Pi_{3/2}$ & 7.06 & 7.05\\
 B$^2\Sigma_{1/2}$ & 4.29 & 4.23\\
\hline \hline
\end{tabular}
\caption{Lifetimes in ns for the electronic excited states of MgF calculated using the simplified r-centroid approximation, Eq.\,\eqref{eq:r-centroid}, and the full $R$-dependent TDM, Eq.\,\eqref{eq:transisiton rate}. }
\label{tab:lifetime}
\end{table}

\section{Experiment}

Experimental data were collected in the same apparatus used in a recent measurement of the decay rates of the the Cr $y^7P^\circ_{2,3,4}$ states \cite{Norrgard2022}.  A cryogenic buffer gas beam (CBGB) of MgF is produced by laser ablation of a sintered MgF$_2$ precursor target.
We orient our experiment by taking  $\hat{z}$ to  be the direction of travel of the molecular beam (roughly horizontal), $\hat{y}$ vertically upward, and $\hat{x}$ parallel to the ground and forming a right-handed coordinate system.

The molecular beam is excited by a nominal 15\,mm 1/e$^2$ diameter laser beam  which is retroreflected such that it   propagates in the $\pm\hat{y}$-directions.  The laser is typically polarized linearly in the $\hat{z}$-direction.   The light is produced using a frequency-doubled titanium-doped sapphire laser.  For decay rate measurements, the light is pulsed on and off using an acousto-optic modulator (AOM). The AOM's first-order diffracted beam is directed to the molecular beam vacuum chamber via a 1\,m-long polarization maintaining optical fiber.  

Fluorescence is collected along the $\pm \hat{x}$ axes by two identical imaging systems.  Each imaging system consists of a broadband antireflection coated fused silica vacuum viewport, a 75 mm focal length plano-convex singlet lens placed one focal length away from the center of the chamber, and a camera lens assembly focused to infinity.  In order to select fluorescence from a particular vibronic transition, one or more interference filters are placed between the singlet lens and camera lens assembly.  To minimize angle-of-incidence-dependent transmission through the filters, the possible angles of incidence on the filters are restricted to $\theta < 7^\circ$ by placing two 18 mm diameter apertures spaced by 150 mm between the singlet lens and the filters.  Photons are detected using photomultipliers (PMTs) and counted using a multichannel event timer with 80\,ps timing resolution.  For fluorescence lifetime measurements, a hybrid PMT-avalanche photodiode is used to remove afterpulsing systmeatic effects \cite{Wahl2020}.  
Timing, computer control of equipment, and data collection are performed using the Labscript Suite  \cite{Starkey2013}.  

 The MgF$_2$ target is ablated using a  10\,ns long pulse of 532\,nm light. Pulse energies between 25~mJ and 50~mJ were found to produce the best yield.
The source is run with a He buffer gas flow rate of $7\times10^5$~Pa~mL/min, or 7 standard cubic centimeters per minute. 
Compared to Cr \cite{Norrgard2022} or sintered SrF$_2$ \cite{Barry2011} precursor targets, the yield from sintered MgF$_2$ precursor targets decays rapidly, dropping by roughly half after typically 100 ablation pulses.
This has motivated us to begin construction of a CBGB based on chemical reaction between Mg metal and a fluorinated gas \cite{Truppe2018,Doppelbauer2022}, which should stably produce high yields.  

\subsection{Branching Fractions}
\label{sec:branching_fractions}
\begin{table*}
\begin{tabular}{lddddd}
\hline\hline
Parameter &  \multicolumn{1}{c}{$b_{00}$} &  \multicolumn{1}{c}{$b_{01}$}&  \multicolumn{1}{c}{$b_{02}$}&  \multicolumn{1}{c}{$b_{03}$}&  \multicolumn{1}{c}{$b_{04^+}$}\\
\hline

 \vspace{6 pt} 
 Branching Fraction - Theory             & 0.968\,7     &0.030\,5     & 0.000\,81   &0.000\,021 & 0.000\,000\,8\\
 Branching Fraction - Measured          & 0.967\,63   & 0.031\,42   & 0.000\,91   & 0.000\,044 & \multicolumn{1}{c}{\quad \quad$< 0.000\,02$} \\
 \vspace{6 pt}Total Uncertainty               & 0.000\,28 & 0.000\,27 & 0.000\,03   & 0.000\,013  \\
 \vspace{2 pt}
 Statistical Uncertainty                   & 0.000\,08   & 0.000\,08   & 0.000\,01   & 0.000\,013  \\

 Systematic Uncertainties: \\
\quad\quad Normalization Discrepancy              & 0.000\,03   & 0.000\,00   & 0.000\,03   & 0.000\,000\\
\quad\quad Calibration - Statistical    & 0.000\,02   & 0.000\,02   & 0.000\,001  & 0.000\,000 \\
\quad\quad Calibration - Power Drift    & 0.000\,12   & 0.000\,12   & 0.000\,004  & 0.000\,000\\
\quad\quad Filter Transmission           & 0.000\,18  & 0.000\,18    & 0.000\,00   & 0.000\,000 \\

Total Systematic Uncertainty 
                             & 0.000\,27 & 0.000\,16 & 0.000\,004  &0.000\,000  \\

\hline\hline
\end{tabular} \caption{Branching Fractions  $b_{0v^{\prime\prime}}$ and $1\sigma$ error budget  for the MgF A$^2\Pi_{1/2}$ state.}
\label{tab:branching fraction}
\end{table*}

In this section we detail the measurement of the branching fractions $b_{0v^{\prime\prime}}$ from the $^{24}$MgF $\ket{A^2\Pi_{1/2}, v^\prime=0;J^\prime=1/2, P^\prime = +}$ state (that is, the excited state typically used in laser cooling applications) to ground vibrational levels $v^{\prime\prime}$ by measuring laser induced fluorescence at transition wavelength $\lambda_{0v^{\prime\prime}}$. 
Table \ref{tab:branching fraction} summarizes the experimental results and  our calculated values derived from weighting the Franck-Condon factors $q_{0v^{\prime\prime}}$ of Section \ref{sec:theory} by the inverse-cube of the transition wavelength $\lambda_{0v^{\prime\prime}}^{-3}$.

Our branching fraction measurement procedure is based on the one used in the Amherst College group's investigation of TlF \cite{Hunter2012, Norrgard2017}. 
Two different bandpass interference filters $i$ and $j$ are inserted into the imaging systems 1 and 2 to simultaneously monitor two vibronic transitions at wavelengths $\lambda_{0i}$ and $\lambda_{0j}$, respectively.
The multichannel event timer stores the detected photon counts from both detectors in a histogram of 2.6\,$\mu$s time bins.
The counts in a 5\,ms interval between ablation pulses are used to determine the mean background counts per bin. 
The background is subtracted from a 1.6\,ms signal interval, which roughly corresponds to the full width at half maximum of the fluorescence signal, to determine the number of fluorescence counts.
The uncertainty in the number of fluorescence counts, listed as ``statistical uncertainty'' in Table \ref{tab:branching fraction}, is determined by the uncorrelated combination of the uncertainty in the signal counts and the uncertainty in the background counts, assuming shot noise.

We observed fluorescence on all vibronic bands up to \mbox{$v^{\prime\prime}=3$}.
An unsuccessful attempt was made to detect fluorescence at wavelength $\lambda_{04}$, which set an experimental upper limit of roughly three times the calculated value for $b_{04}$. 

The fluorescence counts on detector $a$ with filter $i$ is
\begin{equation}
    S^{(a)}_i = N b_{0i} f_i \eta^{(a)}_i,
\end{equation}
where $f_i$ is the transmission efficiency of the filter, $\eta^{(a)}_i$ is the combined geometric and quantum efficiency for detector $a$ at wavelength $\lambda_{0i}$, and $N$ is the total number of molecules excited by the laser.
In each run of the branching fraction experiment, we take the ratio $r_{ij} = S^{(1)}_i / S^{(2)}_j$ of the two signals.  The experiment is then repeated with the filters swapped to obtain the signal ratio $r_{ji}$.
Comparing these two measurements yields the ratio of branching fractions
\begin{equation}
    \frac{b_{0i}}{b_{0j}} = \frac{f_i}{f_j} \sqrt{ \frac{\eta^{(2)}_{j}\eta^{(1)}_{j}}{\eta^{(1)}_{i}\eta^{(2)}_{i}} }\sqrt{\frac{r_{ij}}{r_{ji}}}\,.
\end{equation}
A sufficient number of ratio measurements were taken to over-constrain a fit to the branching fractions $b_{0v^{\prime\prime}}$.  For example, we may derive the ratio $b_{03}/b_{00}$ by either comparing the $i=0,3$  data only 
\begin{equation}
    \frac{b_{03}}{b_{00}} = \frac{f_0}{f_3} \sqrt{ \frac{\eta^{(1)}_{0}\eta^{(2)}_{0}}{\eta^{(1)}_{3}\eta^{(2)}_{3}} }\sqrt{\frac{r_{30}}{r_{03}}}
\end{equation}
or by comparing the  $i=0,1,3$ data
\begin{equation}
         \frac{b_{03}}{b_{00}} =  \frac{f_0}{f_3} \sqrt{\frac{\eta^{(1)}_{0}\eta^{(2)}_{0}}{\eta^{(2)}_{3}\eta^{(1)}_{3}}} \sqrt{\frac{r_{31}}{r_{13}}\frac{r_{10}}{r_{01}}}\,.
\end{equation}
The two methods are statistically consistent for the ratio $b_{03}/b_{00}$, and the weighted mean of these two determinations are used in the analysis.
However, repeating the similar procedure for the ratio $b_{02}/b_{00}$, we find a $3$-$\sigma$ discrepancy between comparisons of the $i=0,2$ data and the $i = 0,1,2$ data.
We take the mean of these two determinations as the reported ratio $b_{02}/b_{00}$, and the difference of the two determinations as the uncertainty in this ratio. 
The impact of this additional uncertainty on the branching fractions $b_{0v^{\prime\prime}}$ is labeled as ``Normalization Discrepancy'' in Table \ref{tab:branching fraction}.

In order to achieve the lowest possible uncertainty branching fraction measurements, it was necessary to accurately characterize the light collection efficiency of our imaging system as well as the transmission properties of the bandpass interference filters.  The response of the entire imaging system was compared to the response of a calibrated photodiode as a function of wavelength in the National Institute of Standards and Technology (NIST) Detector Calibration Facility \cite{Houston2022}.   Spectral responsivity calibration uncertainties as low as 0.005\,\% are possible in this facility.

The Detector Calibration Facility operating principle is to use a quartz-tungsten-halogen lamp filtered by a prism-grating monochromator to provide a low-noise, wavelength-tunable light source.    For our calibration, the 75 mm focal length lens was adjusted to provide a well-collimated beam of light beyond the relay-mirror assembly of the monochromator. In order to reduce the light intensity to below the saturation threshold of the PMTs, two neutral density filters were inserted into the optical beam path of the NIST Detector Calibration Facility.  The transmission of each filter was calibrated separately and the uncertainties are accounted for in the final detection efficiency measurement.  Calibration data were recorded for a few minutes for each setting, until the statistical uncertainty in the response was approximately 0.1\,\%.

  To ensure the calibration was insensitive to the alignment of the light into the imaging system, an iris was placed in front of the first neutral density filter and adjusted so that the beam size at the PMT was smaller than the active detection area of the PMTs. This was confirmed by placing the PMT in a position where there was no change in the observed signal with displacement in either the horizontal and vertical directions.  The optical power was measured with a calibrated photodiode before and after the calibration of each PMT in order to account for drifts in the optical power.

For each detector and filter combination $i\neq 0$, we perform absolute photon collection efficiency calibrations at the the filter's vibronic pass band $\lambda_{0i}$ as well as $\lambda_{00}$.    For filter $i=0$ we calibrated each detector at $\lambda_{00}$ and $\lambda_{01}$. The transmission is  $<10^{-3}$ for  all other wavelength and filter combinations, contributing negligibly to the error estimate. This allows for all molecule fluorescence signals to be corrected for the small amount of $\lambda_{0j}$ ($j\neq i$) light transmitted by filter $i$.  This correction is approximately 10\,\% for ratios involving $v^{\prime\prime}=3$, and less than 1.4\,\% for all stronger transitions.

The dependence of the filter transmission on wavelength and angle of incidence was measured in the NIST facility for Regular Spectral Transmittance \cite{Allen2011}.  The error due to the collection of light at non-zero angles of incidence is estimated by comparing the results of several Zemax \cite{NISTDisclaimer} ray-tracing simulations. The simulations are designed to realistically model the collection optics.
The simulation includes the full optical system less the vacuum viewport.
A 15\,mm diameter, 100\,mm long cylindrical light source (uniform probability for emission of a light ray in both position and direction) oriented perpendicular to the optical collection axis is used to  model the overlap of the excitation laser with the molecular beam. 

From these simulations, the probability for a photon to reach the detector with a given angle of incidence is determined. This calculated collection efficiency is weighted by the measured angle-dependent filter transmission. The angle-dependent transmission uncertainty is estimated as the difference between the calculated transmission for all light at normal incidence and the calculated transmission of our more realistic simulation.  The angle-dependent uncertainty on the transmission is 5\,\% for the $i=3$ filter, which is well below the $30\,\%$ fractional statistical uncertainty for this weak transition.  All other filters were determined to have less than 0.3\,\% angle-dependent uncertainty.  All filters are also assigned a 0.5\,\% transmission uncertainty from their calibration at the NIST facility for Regular Spectral Transmittance.
Simulations with individual optical components displaced by 3\,mm were found to change the calculated transmission by, at most, one order of magnitude less than the estimated angle-dependent transmission uncertainty.

\subsection{Decay Rates}
\label{sec:decay_rates}
\begin{table}
\begin{tabular*}{0.8\columnwidth}{ldd}
\hline\hline
Parameter ($\times 10^{6}\,\rm{s}^{-1}$) &  \multicolumn{1}{c} {A$^2\Pi_{1/2}$} & \multicolumn{1}{c}{A$^2\Pi_{3/2}$}  \\
\hline
 Decay Rate $\Gamma$          &    131.6      &    129.5     \\
 \vspace{6 pt}Total Uncertainty               &      1.4    &      0.9  \\
 
 \vspace{2 pt} Statistical Uncertainty   &      0.9   &      0.8   \\
 Systematic Uncertainties: \\
 \quad Truncation Error          &      0.26 &      0.24  \\
\quad Quantum Beats \\

 \quad\quad Laser Detuning &      1.1   &      0.17  \\
 \quad\quad Laser Polarization &      0.04 &      0.15  \\
 \quad\quad $\mathcal{B}_x$     &     0.03 &      0.15  \\
 \quad\quad $\mathcal{B}_y$     &      0.04 &      0.03 \\
 \quad\quad $\mathcal{B}_z$     &     0.04 &      0.04 \\
 
 \quad Pulse Pileup              &      0.005     &      0.007     \\
 \quad Differential Nonlinearity &      0.014 &      0.014 \\
 \quad Time Calibration          &      0.003     &      0.003     \\
 Total Systematic Uncertainty        &      1.1   &      0.4  \\
 
\hline\hline
\end{tabular*} \caption{Measured decay rates $\Gamma$ and $1\sigma$ error budget  for the MgF A$^2\Pi$ states.}
\label{tab:error budget}
\end{table}

\begin{figure}
    \centering
    \includegraphics[width=\columnwidth]{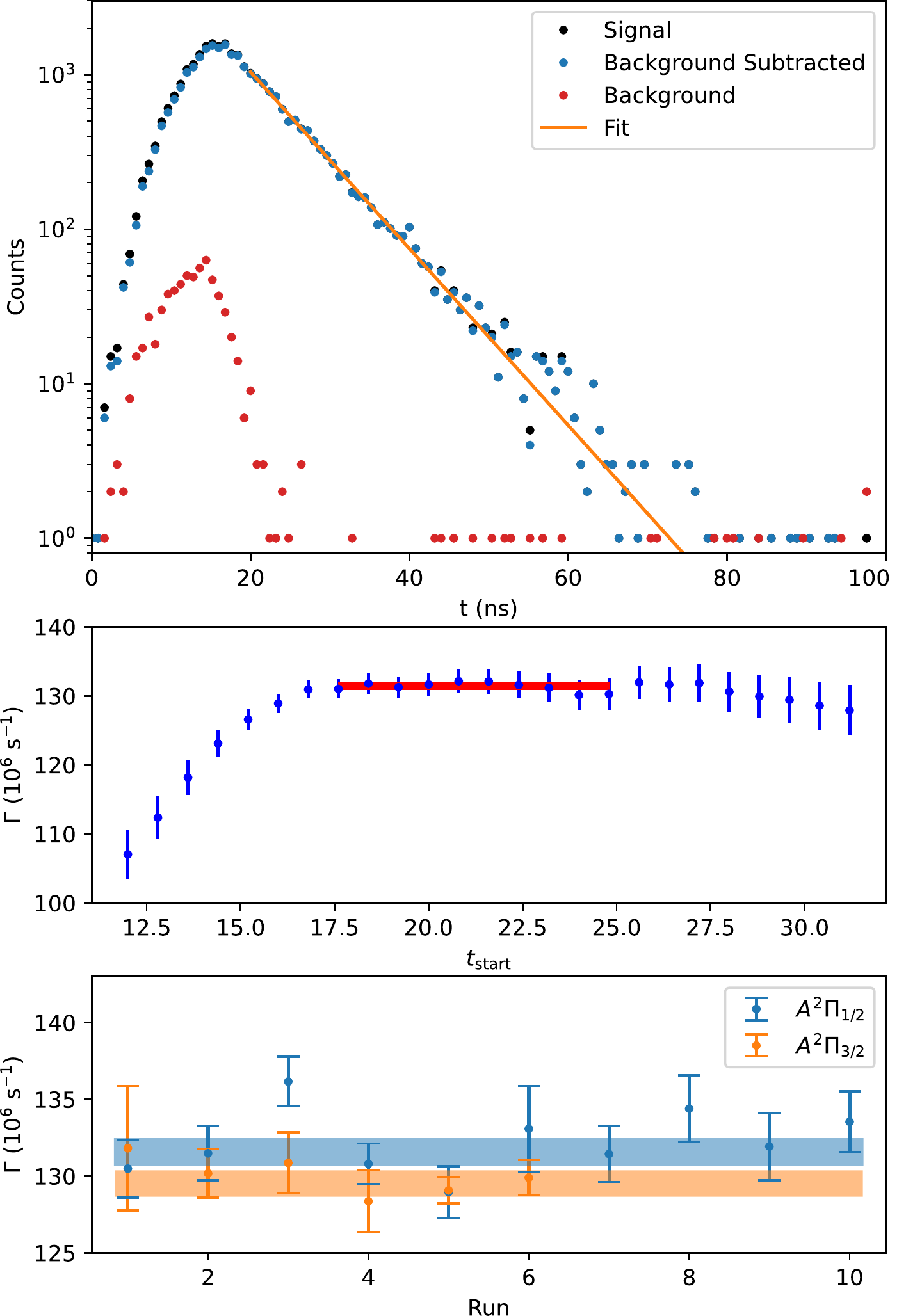}
    \caption{(Top) Histogram of counts as a function of time for a typical run (number 7) to determine decay rate $\Gamma$ for the A$^2\Pi_{1/2}$ state.  The histogram bin width is 800\,ps.  The line labeled ``fit'' is a representative fit to Eq.\,\eqref{eq:single exponential} with $t_{\rm{start}}\,=\,20$\,ns.  (Middle) Fitted $\Gamma$ as a function of $t_{\rm{start}}$ (blue line) for a typical run (number 7).  Error bars denote  the fit uncertainty at each $t_{\rm{start}}$. The range of $t_{\rm{start}}$ chosen for our analysis and the standard deviation of fitted $\Gamma$ over this range are indicated by the range and thickness red line, respectively. (Bottom) Data points fitted $\Gamma$ for each run.  Error bars are the quadrature sum of fit and truncation uncertainties.  Shaded bands denote the $1\sigma$ uncertainty band obtainted by a weighted average of all runs for each excited state.  }
    \label{fig:Lifetime}
\end{figure}

In this section, we detail the measurement and analysis procedures used to determine the spontaneous decay rates $\Gamma$.  Our measured values of $\Gamma$ for the A$^2\Pi_{1/2}$ and A$^2\Pi_{3/2}$  states are presented in Table \ref{tab:error budget} with a list of uncertainty estimates.  Our analysis procedure closely follows that of Ref.\,\cite{Norrgard2022}.

All data for the A$^2\Pi_{1/2}$ state are  collected with the laser  tuned to the P$_1$(1)/Q$_{12}$(1) laser cooling transition.  Because the A$^2\Pi_{3/2}$ state is typically not used for laser cooling, there is no obvious preferred rotational line to investigate. The MgF beam is observed to have the highest population in the $N=3$ rotational state; therefore  all data  for the A$^2\Pi_{3/2}$ state are taken with the laser tuned to the R$_2$(3)/Q$_{21}$(3) line.  Typical laser power for the decay rate measurements is 40\,mW.
To minimize background scattered laser light, fluorescence is monitored on the decay to $v^{\prime\prime}=1$ using the $\lambda_{01} = 368.4$\,nm  interference filter.


In the standard measurement configuration, the AOM produces 1.8$\times 10^4$ light pulses per ablation pulse.  The light pulses have a full width at half maximum of 16\,ns and a pulse repetition rate of 6\,MHz.  The MgF target is ablated at a repetition rate of 2\,Hz.   A histogram of events is collected with 80\,ps time bins by the event timer with data readout performed after 300 ablation pulses.  The procedure is then repeated, absent the ablation pulse, to obtain a background scattered light signal.  This signal plus background measurement procedure is repeated until the desired  1\,\% to 2\,\% statistical uncertainty on the decay rate is achieved, with the cummulativie data constituting an experimenatl ``run''.   A single run lasts roughly 40\,min. We performed 10 runs measuring the decay rate of the A$^2\Pi_{1/2}$ and 6 runs measuring the A$^2\Pi_{3/2}$.  Our reported value for the decay rates in Table \ref{tab:error budget} are the weighted means of all runs for each state.  In addition, several runs were performed on each state with one experimental parameter varied in order to quantify potential systematic errors.


\subsubsection{Pulse Pileup}
We account for  missed counts due to coincident events, or ``pulse pileup'', by following the procedure outlined in Ref.\,\cite{Norrgard2022}.
Decay rate data are typically collected at a rate of roughly 1 count per 1000 excitation cycles.  The multihit-capable event timer has a dead time of 650\,ps, and the photodetector has a maximum average count rate of 80\,MHz. We correct the raw signal counts $N_i$  in each time bin $i$ to obtain $N_i^\prime$ corrected counts per time bin using
\begin{equation}\label{eq:pulse pileup}
    N_i^\prime = \frac{N_i}{1-\frac{1}{N_{\rm{cycle}}}\sum_{j=i-k_d}^i N_j}
\end{equation}
where $N_{\rm{cylcle}}$ is the total number of excitation cycles, and $k_d = t_{\rm{bin}}\times 80\,$MHz is the number of time bins of width $ t_{\rm{bin}}$ in the detector-limited deadtime \cite{Patting2007}.  We estimate the error $\delta\Gamma$ due to this correction by analyzing the data assuming the detector has no response after the first detection of an excitation cycle.  The difference  is less than $5\times 10^{-5}$ fractional uncertainty $\delta\Gamma/\Gamma$ for our measurements.

\subsubsection{Fit Procedure and Truncation Error}
After correcting for pulse pileup, background counts $N_i^{\rm{bg}\prime}$ are subtracted from the signal counts $N_i^{\rm{sig}\prime}$.  The resulting data are binned in 800\,ps intervals and fit to a single decaying exponential
\begin{equation}\label{eq:single exponential}
    N_i^{\rm{sig}\prime}-N_i^{\rm{bg}\prime} = Ae^{-\Gamma t_i}.
\end{equation}
When a constant offset was added to Eq.\,\eqref{eq:single exponential}, the offset was found to be statistically consistent with zero and did not change the fitted decay rate within its  statistical uncertainty.
Pulse-pileup-corrected data for a typical experimental run measuring the A$^2\Pi_{1/2}$ decay rate are shown in upper panel of Fig.\,\ref{fig:Lifetime}.

The arbitrary choice of the start time $t_{\rm{start}}$ of the fit may influence the decay rate measurement (sometimes called truncation error).   
We vary $t_{\rm{start}}$, setting this parameter to the start of each $t_{\rm{bin}}=800$\,ps time bin over a 7\,ns range which corresponds to between about 50\,\% to 20\,\% of the peak observed counts (Fig.\,\ref{fig:Lifetime} middle panel). 
For each run, we assign a value for $\Gamma$ by taking the average fitted $\Gamma$ over this 7\,ns range of start times, weighted by the nonlinear least-squares $1\sigma$ confidence interval when fitting to Eq.\,\eqref{eq:single exponential}.  We assign a statistical uncertainty equal to the median $1\sigma$ nonlinear least-squares fit uncertainty in the  range, and assign a truncation uncertainty equal to the standard deviation of the fitted $\Gamma$ values. 

The assigned decay rate $\Gamma$ and the combined statistical and truncation uncertainties for each run are shown in the bottom panel of Fig.\,\ref{fig:Lifetime}.  The shaded bands depict the weighted average and standard error of all runs for each state.  The calculations of Section \ref{sec:theory} predict the A$^2\Pi_{3/2}$ decay rate to be $6\times10^5$\,s$^{-1}$ faster than A$^2\Pi_{1/2}$  decay rate, that is,  within the statistical sensitivity of our measurement.  Indeed, accounting for all sources of uncertaintity detailed below, we find no statistically significant difference in the decay rates MgF A$^2\Pi_{1/2}$ \mbox{($\Gamma = 1.316(14)\times10^8$\,s$^{-1}$)} and A$^2\Pi_{3/2}$ \mbox{($\Gamma = 1.295(9)\times10^8$\,s$^{-1}$)} states.

\subsubsection{Quantum Beats}

In addition to the exponential decay expected in Eq.\,\eqref{eq:single exponential}, an oscillatory amplitude may be observed due to hyperfine and Zeeman quantum beats.
The laser pulse has a full width at half maximum of 16\,ns, corresponding to a Fourier limited linewidth of about $2\pi\times 100$\,MHz.
For the A$^2\Pi_{1/2}$ decay rate measurement, the laser is typically tuned to the peak in fluorescence around the unresolved $\ket{X^2\Sigma^+ ; N=1,J=3/2,F = 1,2, P=-} \rightarrow 
 \ket{A^2\Pi_{1/2}; J^\prime=1/2, F^\prime = 0,1,  P^\prime=+}$
transitions.  For these transitions, electric dipole selection rules restrict all ground states to being excited to one or the other excited state $F^\prime = 0,1$ manifold, but not both.  Therefore, hyperfine quantum beats should not be observed in the A$^2\Pi_{1/2}$. For the A$^2\Pi_{3/2}$ decay rate measurement, the partially resolved $\ket{X^2\Sigma^+ ; N=3,J=7/2,F = 3,4, P=-} \rightarrow 
 \ket{A^2\Pi_{3/2}; J^\prime=7/2, F^\prime = 3,4,  P^\prime=+}$
transitions are excited. The $F^\prime=3,4$ levels are separated in energy by about $\Delta E/\hbar = 2\pi\times 31$\,MHz \cite{Doppelbauer2022}, and therefore we may expect to see evidence of hyperfine quantum beats in the  R$_2$(3)/Q$_{21}$(3)  signal.  

We test for systematic errors due to hyperfine quantum beats by attempting to preferentially excite of one or the other excited state hyperfine level by varying the laser detuning by $+20\,$MHz and $-20\,$MHz  from the fluorescence peak. We found a statistical difference in the fitted $\Gamma$ when detuning the laser for the A$^2\Pi_{1/2}$ level; the uncertainty obtained by linear regression is  $\delta\Gamma/\Gamma = 0.8$\,\% for this level.  As this was within our target accuracy, the effect was not investigated further.  The uncertainty due to laser detuning on the A$^2\Pi_{3/2}$ decay rate is roughly one order of magnitude smaller.

Possible systematic error due Zeeman quantum beats is substantially mitigated compared to our recent Cr decay rate measurements \cite{Norrgard2022} by improved control of the magnetic field.  Three pairs of Helmholtz coils surround the interaction region of the experiment.  Using the Helmholtz coils and a three-axis Hall sensor, we map out the small field hysteresis loop of the vacuum chamber such that we may set the residual magnetization in a deterministic way.  We then apply a bias field to either null the residual magnetization for standard data, or apply a known magnetic field of up to 0.2\,mT along each axis to exaggerate the effects of uncancelled magnetic fields.  We estimate the uncertainty in the nulled magnetic field to be 8\,$\mu$T.

For the A$^2\Pi_{1/2}$ state, the $g$-factor is expected to be small ($g <0.01$), and thus Zeeman quantum beats should be negligible even for the largest 0.2\,mT magnetic fields applied.  
Indeed, no statistically significant magnetic field effect on $\Gamma$ is observed for this state. 
Regardless of the magnitude of the $g$-factor, our experimental geometry (excitation light polarized along $\hat{z}$ and imaging along $\hat{x}$), should preclude observing Zeeman quantum beats for applied $\mathcal{B}_x$ or $\mathcal{B}_z$.  
We observed no statistically significant effect in these cases for the A$^2\Pi_{3/2}$, though a relatively poor constraint on the effect of uncancelled $\mathcal{B}_x$ was obtained due to a smaller number of measurements taken. It should be possible to observe
Zeeman quantum beats for an applied $\mathcal{B}_y$ field, and 
an attempt was made to assign a $g$ factor to the A$^2\Pi_{3/2}$ state by applying a large bias field $\mathcal{B}_y=2$\,mT. While a the fitted $\Gamma$ under these conditions increased to $1.38(3)\times 10^8$\,s$^{-1}$, the contrast was too poor to reliably fit an oscillatory term and assign a $g$-factor.  For all cases, the fractional uncertainty due to uncanceld magnetic fields is $\delta\Gamma/\Gamma < 0.12$\,\%

Finally, we tested for the effect of laser polarization on possible quantum beats by exciting with light linearly polarized along $\hat{x}$ instead of along $\hat{z}$.  For zero magnetic field, we expect the laser polarization to have no effect. A statistically significant deviation in the decay rate was observed for the A$^2\Pi_{3/2}$ level leading to a fractional uncertainty of $\delta\Gamma/\Gamma = 0.15$\,\%.  Again, this effect was sufficiently small that it was not further investigated.  No statistically significant shift was observed for the A$^2\Pi_{1/2}$ level.

\subsubsection{Time Calibration}
Uncertainties  due to differential nonlinearity (that is, non-cumulative uncertainty in the bin width of the event timer) and time calibration of the event timer are assigned in an identical manner to that described in our Cr decay rate measurement \cite{Norrgard2022}.  The uncertainty due to these effects is at least several times smaller than the other sources of uncertainty considered above.


\section{Comparison of results and conclusion}
\label{sec:conclusion}
Table \ref{tab:decay rate comparison} compares the experimental and calculated decay rates of the A$^2\Pi$ state determined in this work with values reported elsewhere.  

By using a calibrated imaging system, we have determined the vibrational branching fractions of MgF to an unprecedented accuracy.  This sets a precision benchmark for vibrational branching calculations, and we find excellent agreement  with values calculated by our multireference relativistic Fock-space coupled cluster method (within 0.1\,\% for $b_{00}$).
Given this near-perfect match as well as an exceptional agreement in transition energies (theoretical $T_e$ values agree within 0.1\,\% with experiment, Table \ref{tab:spectroscopic}), the observed overshooting of theoretical decay rates (Table \ref{tab:decay rate comparison}) can be almost entirely attributed to the uncertainty in the calculated TDM values. 

\begin{figure}
    \centering
    \includegraphics[width=\columnwidth]{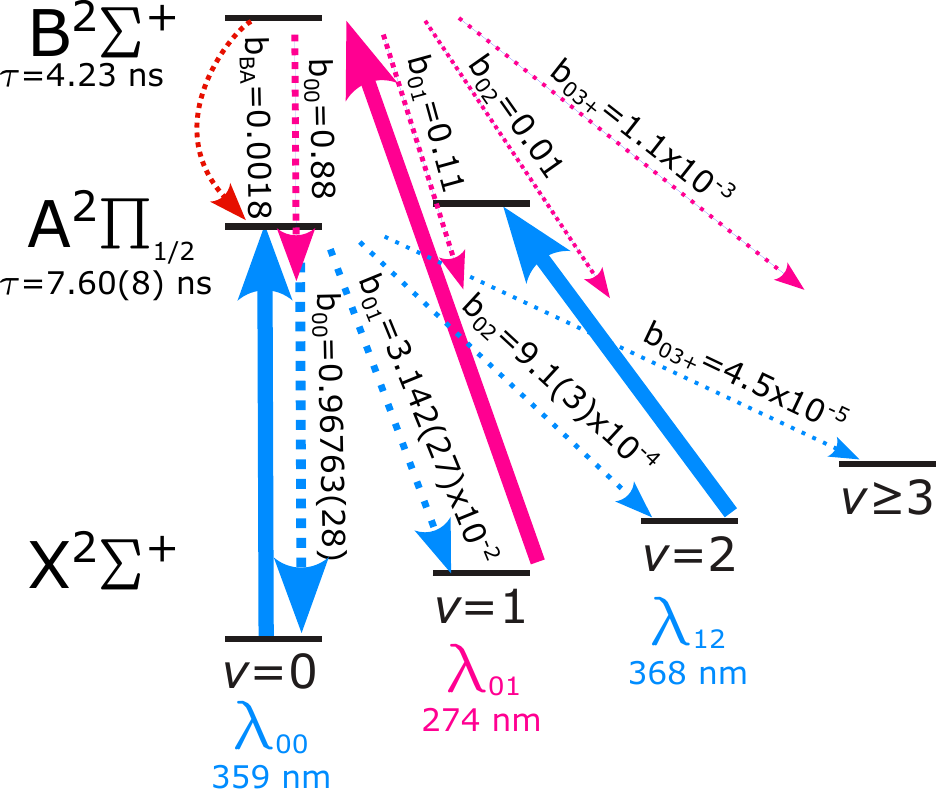}
    \caption{Possible optical cycling scheme for MgF.  Blue, pink, and red arrows denote X-A, X-B, and A-B transitions, respectively.  Solid lines denote laser excitation.  Dashed lines denote radiative decay. Values without uncertainties are calculated.}
    \label{fig:cycling}
\end{figure}

The radiative decay rate and vibrational branching fractions for the MgF A$^2\Pi_{1/2}$ state are highly favorable for laser cooling.
With three lasers, it should be possible to scatter up to $1/b_{03} \approx 2\times  10^4$ photons.  One possible optical cycling scheme is depicted Fig.\,\ref{fig:cycling}.  By repumping the $v^{\prime\prime}=1$ level through the B$^2\Sigma^+$  state instead of the A$^2\Pi_{1/2}$ state, this scheme avoids producing a $\Lambda$ system between the main cycling laser (wavelength $\lambda_{00}$) and first repump laser (wavelength $\lambda_{01}$) \cite{NorrgardThesis}.  When driving P$_1$(1) transitions with this scheme, the maximum possible scattering rate is $R_{\rm{sc}}^{\rm{max}} = 0.25 \Gamma$.   This comes at the cost of an additional loss channel due to B$\rightarrow$A decays, with roughly $6\times 10^{-5}$ probability per optical cycle.  Molecules will subsequently decay to the $\ket{\text{X}^2\Sigma^+, v^{\prime\prime}=0; N^{\prime\prime}=0,2}$ levels.  The states could be repumped by driving $N^{\prime\prime}=0 \leftrightarrow N^{\prime\prime}=1\leftrightarrow N^{\prime\prime}=2$ microwave transitions, or by a combination of $N^{\prime\prime}=0 \leftrightarrow N^{\prime\prime}=1$ microwaves and an $N^{\prime\prime}=2$ repump laser \cite{Norrgard2016}.

While the maximum possible scattering rate for the scheme depicted in Fig.\,\ref{fig:cycling} is $R_{\rm{sc}}^{\rm{max}} = 0.25 \Gamma$, using the pyLCP python package \cite{Eckel2022} to simulate a magneto-optical trap (MOT) with realistic parameters, we find that the maximum rate of decelerating photon scatters is typically closer to $R_{\rm{sc}}\approx 0.05 \Gamma$.  Nonetheless, this photon scattering rate is sufficient to stop molecules with initial velocities up to 100\,m/s in a distance of only 3\,cm.  This small stopping distance suggests it may be possible to directly load a MgF MOT from a CBGB source without additional slowing mechanisms \cite{Hemmerling2014}, which would enable the loading of multiple molecule pulses from the CBGB \cite{Shaw2020}.  Details of these simulations will be explored in a future work \cite{Rodriguez2023}.

\begin{table}[t]
\begin{tabular}{cccc}
\hline\hline
Method & \multicolumn{2}{c}{Decay Rate $\Gamma (\times 10^6$\,s$^{-1}$)} & Reference \\
\cmidrule(lr){2-3}

&  {A$^2\Pi_{1/2}$} & \multicolumn{1}{c}{A$^2\Pi_{3/2}$}  \\
\hline \\ 
 Experiment                     & 131.6(1.4)     & 129.5(9)      &this work   \\
 \vspace{6 pt} Theory & 141.3 & 141.9& this work\\
 Experiment & 138(7) & -- & \cite{Doppelbauer2022}\\
Theory & \multicolumn{2}{c}{--~~~~~139.6~~~~--~~~} & \cite{Pelegrini2005}\\
Theory & \multicolumn{2}{c}{--~~~~~125.6~~~~--~~~} & \cite{Kang2015}\\
\hline\hline
\end{tabular} \caption{Comparison of our measured and calculated decay rates to values from previous works. Values in parentheses are the combined $1\sigma$ statistical and systematic uncertainty.}\label{tab:decay rate comparison}
\end{table}

\begin{acknowledgments}
The authors thank  National Institute of Standards and Technology (NIST) colleagues Zeeshan Ahmed, Daniel Barker, Joe Rice, Julia Scherschligt, Ian Spielman, and Joseph Tan for providing equipment used in these experiments. The authors thank Thinh Bui, Jacob Higgins, and Eric Shirley for comments on the manuscript. 
 Financial support was provided by NIST. The authors thank the Center for Information Technology of the University of Groningen for their support and for providing access to the Peregrine high-performance computing cluster.

\end{acknowledgments}

\bibliography{thebib}

\begin{thebibliography}{50}
\bibitem [{NIS()}]{NISTDisclaimer}%
  \BibitemOpen
  \href@noop {} {\bibinfo {title} {{Any mention of commercial products within
  this work is for information only; it does not imply recommendation or
  endorsement by NIST}}}\BibitemShut {NoStop}%
\end{thebibliography}

\section{Supplementary Information}

\subsection{Dipole moments}
We compare the PDMs calculated with the MRCI and the FSCC methods. Furthermore, to validate the switch from one computational toolbox to another, we compare the TDMs and PDMs calculated with the DIRAC and Molpro programs (using the MRCI approach). In all the cases, we use the d-aug-dyall.ae4z basis set, and we freeze 4 electrons. 
In the calculations carried out with MRCI and FSCC methods using the DIRAC program, we included in the correlation description all the virtual orbitals up to 5 and 10 a.u, respectively, and we used a 2-component relativistic formalism (X2C). In FSCC sector (0,1), the Mg $3s$ and $3p$ orbitals were included in the active particle space. In MRCI, an analogous CASSCF reference was used with the complete active space usually designated as (1,4), i.e. 1 electron in 4 MOs. In Molpro MRCI calculation, we used a similar setup, except that
all the virtual orbitals were correlated and the spin-orbit (SO) contribution was calculated \textit{a posteriori} following a scalar relativistic MRCI calculation. The MRCI results obtained using the DIRAC and Molpro programs shown in Table~\ref{tab:PDM} agree very well so that there is no significant effect from the reduced virtual active space or the different treatment of SO effects. 
For the PDMs, we observe a good agreement between the two MRCI approaches as well as between MRCI and FSCC. The PDM of the excited B$^2\Sigma_{1/2}$ state is more sensitive to the basis set size and the virtual cutoff compared to the lower states. However, this state is not the focus of the present study.
Overall, the calculated values agree very well with the experimental ones.

\begin{table}[h]
\centering
\begin{tabular}{ccccc}
\hline
\multirow{2}{*}{Transition/State}  & MRCI & MRCI & FSCC & \multirow{2}{*}{Expt.} \\
& Molpro & DIRAC & DIRAC &   \\
\hline
 A$^2\Pi_{1/2}$ -- X$^2\Sigma_{1/2}$    &1.837 & 1.798 &      &      \\
 A$^2\Pi_{3/2}$ -- X$^2\Sigma_{1/2}$    &1.837 & 1.798 &      &      \\
 B$^2\Sigma_{1/2}$ -- X$^2\Sigma_{1/2}$ &1.499 & 1.507 &      &      \\
 A$^2\Pi_{3/2}$ -- A$^2\Pi_{1/2}$       &0.001 & 0.003 &      &      \\
 B$^2\Sigma_{1/2}$ -- A$^2\Pi_{1/2}$    &0.332 & 0.310 &      &      \\
 B$^2\Sigma_{1/2}$ -- A$^2\Pi_{3/2}$    &0.332      & 0.310 &      &      \\
 \hline
 X$^2\Sigma_{1/2}$  & 1.220 & 1.210& 1.233 & 1.13(8) \\
 A$^2\Pi_{1/2}$     & 1.398 & 1.365& 1.399 & 1.26(9) \\
 A$^2\Pi_{3/2}$     & 1.398 & 1.362& 1.396 & 1.26(9) \\
 B$^2\Sigma_{1/2}$  & 0.265 & 0.270& 0.225 &      \\  
 \hline \hline      
\end{tabular}
\caption{Transition (top) and permanent (below) dipole moments in 
$e a_0$ at the ground state bond length calculated using the MRCI and FSCC methods. Experimental values obtained from reference \cite{Doppelbauer2022}.}
\label{tab:PDM}
\end{table}

\subsection{Franck--Condon factors}

Table \ref{tab:FCF} presents the FCFs for the vibronic transitions between the ground and excited electronic states of the MgF molecule. We used the potential energy surfaces obtained at the 4c-FSCC(0,1) level of theory and the d-aug-dyall.ae4z basis set, and extracted the FCFs from the LEVEL16\cite{Roy17} program. 

\begin{table*}[] 
\centering 
\begin{tabular}{lllllll} 
\hline \hline 
\multicolumn{7}{c}{FCFs}\\ & $v''=0$ & $v''=1$ & $v''=2$ & $v''=3$ & $v''=4$ & $v''=5$ \\ 
\hline 
\diagbox[]{A$^2\Pi_{1/2}$}{X$^2\Sigma_{1/2}$}& \multicolumn{6}{c}{} \\ 
$v’=0$ & $9.68\times10^{-1}$ & $3.08\times10^{-2}$ & $8.61\times10^{-4}$ & $2.61\times10^{-5}$ & $9.28\times10^{-7}$ & $4.04\times10^{-8}$ \\ $v’=1$ & $3.14\times10^{-2}$ & $9.06\times10^{-1}$ & $5.97\times10^{-2}$ & $2.56\times10^{-3}$ & $1.05\times10^{-4}$ & $4.66\times10^{-6}$ \\ $v’=2$ & $2.40\times10^{-4}$ & $6.23\times10^{-2}$ & $8.45\times10^{-1}$ & $8.67\times10^{-2}$ & $5.04\times10^{-3}$ & $2.60\times10^{-4}$ \\ $v’=3$ & $4.92\times10^{-7}$ & $7.32\times10^{-4}$ & $9.25\times10^{-2}$ & $7.86\times10^{-1}$ & $1.12\times10^{-1}$ & $8.28\times10^{-3}$ \\ $v’=4$ & $4.53\times10^{-13}$ & $2.06\times10^{-6}$ & $1.49\times10^{-3}$ & $1.22\times10^{-1}$ & $7.28\times10^{-1}$ & $1.35\times10^{-1}$ \\ $v’=5$ & $2.03\times10^{-12}$ & $2.09\times10^{-11}$ & $5.33\times10^{-6}$ & $2.55\times10^{-3}$ & $1.51\times10^{-1}$ & $6.72\times10^{-1}$ \\ 
\hline 
\diagbox[]{A$^2\Pi_{3/2}$}{X$^2\Sigma_{1/2}$}& \multicolumn{6}{c}{} \\ 
$v’=0$ & $9.68\times10^{-1}$ & $3.12\times10^{-2}$ & $8.81\times10^{-4}$ & $2.69\times10^{-5}$ & $9.63\times10^{-7}$ & $4.20\times10^{-8}$ \\ $v’=1$ & $3.18\times10^{-2}$ & $9.05\times10^{-1}$ & $6.04\times10^{-2}$ & $2.61\times10^{-3}$ & $1.08\times10^{-4}$ & $4.84\times10^{-6}$ \\ $v’=2$ & $2.47\times10^{-4}$ & $6.31\times10^{-2}$ & $8.44\times10^{-1}$ & $8.77\times10^{-2}$ & $5.15\times10^{-3}$ & $2.68\times10^{-4}$ \\ $v’=3$ & $5.17\times10^{-7}$ & $7.53\times10^{-4}$ & $9.36\times10^{-2}$ & $7.84\times10^{-1}$ & $1.13\times10^{-1}$ & $8.45\times10^{-3}$ \\ $v’=4$ & $1.60\times10^{-12}$ & $2.17\times10^{-6}$ & $1.54\times10^{-3}$ & $1.23\times10^{-1}$ & $7.25\times10^{-1}$ & $1.36\times10^{-1}$ \\ $v’=5$ & $1.97\times10^{-12}$ & $3.57\times10^{-11}$ & $5.60\times10^{-6}$ & $2.62\times10^{-3}$ & $1.52\times10^{-1}$ & $6.69\times10^{-1}$ \\ 
\hline \diagbox[]{B$^2\Sigma_{1/2}$}{X$^2\Sigma_{1/2}$}& \multicolumn{6}{c}{} \\ 
$v’=0$ & $8.80\times10^{-1}$ & $1.09\times10^{-1}$ & $1.01\times10^{-2}$ & $8.58\times10^{-4}$ & $7.12\times10^{-5}$ & $5.98\times10^{-6}$ \\ $v’=1$ & $1.15\times10^{-1}$ & $6.72\times10^{-1}$ & $1.83\times10^{-1}$ & $2.64\times10^{-2}$ & $3.04\times10^{-3}$ & $3.19\times10^{-4}$ \\ $v’=2$ & $4.44\times10^{-3}$ & $2.07\times10^{-1}$ & $5.05\times10^{-1}$ & $2.30\times10^{-1}$ & $4.57\times10^{-2}$ & $6.71\times10^{-3}$ \\ $v’=3$ & $4.73\times10^{-5}$ & $1.23\times10^{-2}$ & $2.78\times10^{-1}$ & $3.73\times10^{-1}$ & $2.56\times10^{-1}$ & $6.60\times10^{-2}$ \\ $v’=4$ & $1.52\times10^{-8}$ & $1.70\times10^{-4}$ & $2.29\times10^{-2}$ & $3.33\times10^{-1}$ & $2.70\times10^{-1}$ & $2.66\times10^{-1}$ \\ $v’=5$ & $1.81\times10^{-9}$ & $3.58\times10^{-8}$ & $3.80\times10^{-4}$ & $3.54\times10^{-2}$ & $3.75\times10^{-1}$ & $1.90\times10^{-1}$ \\ 
\hline \diagbox[]{A$^2\Pi_{3/2}$}{A$^2\Pi_{1/2}$}& \multicolumn{6}{c}{} \\
$v’=0$ & $1.00\times10^{0}$ & $1.50\times10^{-6}$ & $4.08\times10^{-9}$ & $2.95\times10^{-11}$ & $3.87\times10^{-13}$ & $8.08\times10^{-15}$ \\ $v’=1$ & $1.50\times10^{-6}$ & $1.00\times10^{0}$ & $3.00\times10^{-6}$ & $1.24\times10^{-8}$ & $1.21\times10^{-10}$ & $1.85\times10^{-12}$ \\ $v’=2$ & $3.81\times10^{-9}$ & $3.00\times10^{-6}$ & $1.00\times10^{0}$ & $4.50\times10^{-6}$ & $2.49\times10^{-8}$ & $3.01\times10^{-10}$ \\ $v’=3$ & $2.67\times10^{-11}$ & $1.16\times10^{-8}$ & $4.50\times10^{-6}$ & $1.00\times10^{0}$ & $6.01\times10^{-6}$ & $4.15\times10^{-8}$ \\ $v’=4$ & $3.44\times10^{-13}$ & $1.09\times10^{-10}$ & $2.33\times10^{-8}$ & $6.01\times10^{-6}$ & $1.00\times10^{0}$ & $7.53\times10^{-6}$ \\ $v’=5$ & $7.14\times10^{-15}$ & $1.64\times10^{-12}$ & $2.72\times10^{-10}$ & $3.89\times10^{-8}$ & $7.53\times10^{-6}$ & $1.00\times10^{0}$ \\ 
\hline \diagbox[]{B$^2\Sigma_{1/2}$}{A$^2\Pi_{1/2}$}& \multicolumn{6}{c}{} \\
$v’=0$ & $9.69\times10^{-1}$ & $3.03\times10^{-2}$ & $1.15\times10^{-3}$ & $4.51\times10^{-5}$ & $1.89\times10^{-6}$ & $8.27\times10^{-8}$ \\ $v’=1$ & $3.14\times10^{-2}$ & $9.10\times10^{-1}$ & $5.56\times10^{-2}$ & $3.20\times10^{-3}$ & $1.68\times10^{-4}$ & $8.75\times10^{-6}$ \\ $v’=2$ & $8.73\times10^{-5}$ & $5.99\times10^{-2}$ & $8.57\times10^{-1}$ & $7.65\times10^{-2}$ & $5.91\times10^{-3}$ & $3.88\times10^{-4}$ \\ $v’=3$ & $7.95\times10^{-7}$ & $2.15\times10^{-4}$ & $8.57\times10^{-2}$ & $8.11\times10^{-1}$ & $9.36\times10^{-2}$ & $9.10\times10^{-3}$ \\ $v’=4$ & $3.39\times10^{-9}$ & $3.69\times10^{-6}$ & $3.51\times10^{-4}$ & $1.09\times10^{-1}$ & $7.70\times10^{-1}$ & $1.07\times10^{-1}$ \\ $v’=5$ & $1.03\times10^{-10}$ & $1.12\times10^{-8}$ & $1.06\times10^{-5}$ & $4.65\times10^{-4}$ & $1.30\times10^{-1}$ & $7.34\times10^{-1}$ \\ 
\hline \diagbox[]{B$^2\Sigma_{1/2}$}{A$^2\Pi_{3/2}$}& \multicolumn{6}{c}{} \\
$v’=0$ & $9.69\times10^{-1}$ & $2.99\times10^{-2}$ & $1.13\times10^{-3}$ & $4.40\times10^{-5}$ & $1.83\times10^{-6}$ & $7.97\times10^{-8}$ \\ $v’=1$ & $3.10\times10^{-2}$ & $9.11\times10^{-1}$ & $5.49\times10^{-2}$ & $3.14\times10^{-3}$ & $1.64\times10^{-4}$ & $8.48\times10^{-6}$ \\ $v’=2$ & $8.32\times10^{-5}$ & $5.91\times10^{-2}$ & $8.59\times10^{-1}$ & $7.56\times10^{-2}$ & $5.80\times10^{-3}$ & $3.78\times10^{-4}$ \\ $v’=3$ & $8.03\times10^{-7}$ & $2.04\times10^{-4}$ & $8.46\times10^{-2}$ & $8.13\times10^{-1}$ & $9.25\times10^{-2}$ & $8.93\times10^{-3}$ \\ $v’=4$ & $3.14\times10^{-9}$ & $3.72\times10^{-6}$ & $3.31\times10^{-4}$ & $1.08\times10^{-1}$ & $7.72\times10^{-1}$ & $1.06\times10^{-1}$ \\ $v’=5$ & $1.02\times10^{-10}$ & $1.02\times10^{-8}$ & $1.06\times10^{-5}$ & $4.37\times10^{-4}$ & $1.29\times10^{-1}$ & $7.37\times10^{-1}$\\ 
\hline 
\end{tabular} 
\caption{FCFs for the vibronic transitions between the upper ($v'$) and lower ($v''$) electronic states.} 
\label{tab:FCF} 
\end{table*}

\end{document}